\documentclass[aip,jcp,reprint]{revtex4-2}

\usepackage{amsmath}
\usepackage{bm}
\usepackage{graphicx}

\begin{document}


\title{Uniaxial Ordering 
by Self-Assembly of Isotropic Octahedral Junctions} 



\author{Kazuya Saito}
\email{saito.kazuya.gm@u.tsukuba.ac.jp}
\affiliation{Center for Computational Sciences, University of Tsukuba,\\
Tsukuba, Ibaraki 305-8577, Japan}
\affiliation{Research Center for Thermal and Entropic Science,
Graduate School of Science,\\ Osaka University, Toyonaka, Osaka 560-0043, Japan}


\date{\today}

\begin{abstract}
We demonstrate that isotropic octahedral (sixfold branched) junctions 
with three diagonal endpoint pairs of different colors almost inevitably form 
a macroscopic assembly of uniaxial order, exhibiting the perfect order of a single color. 
Monte Carlo simulations of the antiferromagnetic three-state Potts model 
on the tripartite \textbf{\textit{reo}} net, 
consisting of corner-sharing regular octahedrons, confirm this counterintuitive prediction,
while showcasing switching self-assembly upon an ordering phase transition. 
The possible inequivalence of three directions, i.e., more symmetry breaking than uniaxiality, 
is found and discussed for the ordered phase of this model at finite temperatures. 
Some additional analyses of the model are provided, 
including the possibility of a metastable isotropic order, which aligns better with intuition.
\end{abstract}

\maketitle 

\section{Introduction}

Symmetry breaking \cite{nambu} is an essential concept 
in the contemporary understanding of the world. 
The broken symmetry of a macroscopic ensemble is 
usually consistent with the symmetry of the constituting entities. 
Indeed, the uniaxial symmetry of the nematic liquid crystals reflects 
the marked anisotropy of molecules (after averaging molecular details).\cite{onsager,MS} 
This situation also applies to more abstract spin systems, 
as exemplified by the uniaxial magnetization and the directional property of spins.\cite{ising1,ising2} 
In this paper, 
we present a counterintuitive example of uniaxial symmetry brought about 
by the self-assembly of isotropic entities.
 
\begin{figure}[tb]
\begin{center}
\includegraphics[width=8cm]{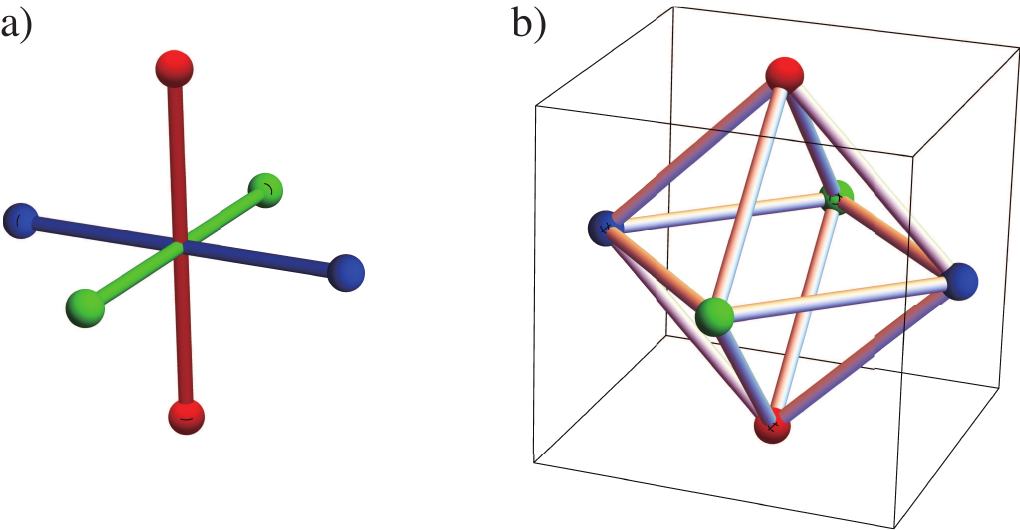}
\end{center}
\caption{a) Sixfold branched junction of three orthogonal segments of three colors.
b) A unit cell of the \textbf{\textit{reo}} net, containing an octahedron.
Spins are located at the spheres on the square faces of a cubic cell (not at the corners).
Spins at sites of different colors belong to different sublattices.}
\label{reo}
\end{figure}

Our system is an ensemble of octahedral (sixfold branched) junctions 
comprising three segments in different colors:
red (R), green (G), and blue (B), as illustrated in Fig.\ \ref{reo}a.
The junction is isotropic since the three pairs are equivalent to each other. 
If we set the rule that the endpoints can only connect between the same colors, 
this ensemble orders uniaxially (as in Fig.\ \ref{junglegym}a) 
with an overwhelming probability than the isotropic order. 
We will probe this later.
 
We will also study this ordering process computationally. 
To treat the problem from an ordinary spin perspective, 
we adopt the antiferroic three-state Potts model\cite{potts,3Potts} 
on the lattice of corner-sharing octahedra (Fig.\ \ref{reo}b). 
The model disfavors the same states (colors) for the interacting spins. 
Since the spin configuration on the regular octahedron involves 
three diagonal pairs of the same state in the minimum energy state, 
those on the lattice are equivalent to 
the ensemble of the octahedral junctions in the connected state.
Note that this study does not intend 
to fully characterize the spin model in the present setting numerically,
although the system has been intact in the past within the author's knowledge.

The lattice of corner-sharing octahedra is known as the \textbf{\textit{reo}} net 
in the reticular chemical context.\cite{rcsr,netCub} 
It is tripartite, by which the property is meant 
to be just complete three-color labeling of all sites with different colors for neighboring sites. 
When the number of sites on each sublattice is the same, 
the net is termed ``balanced.''
The \textbf{\textit{reo}} net is balanced.
It consists of three partially overlapping sublattices,
each of which is a simple cubic lattice of a respective color.
The triangle as the minimum closed interaction path is essential 
for the antiferroic three-state Potts model 
because three colors are necessary. 
The number of balanced tripartite nets with sufficiently high symmetry, 
i.e., containing only triangles as a small closed path, are known to be four: 
the octahedron in dimension three, the icositetrachoron in dimension four, 
the triangular net in dimension two, and the icositetrachoron honeycomb in dimension four.\cite{polyhedron,cluster,exMS_4d}
While the first two are generalized regular polyhedra (polytopes), 
the last two are infinitely extendable over the space in their respective dimensions. 
The stable spin configuration on these nets is unique except for the permutation of three colors. 
In contrast to these, the \textbf{\textit{reo}} net, though it is of relatively high symmetry, 
contains triangles and squares. 
Still, each site has only connections with sites of the sublattices to whom the site does not belong.
Since the coloring is mostly complete within an octahedron with six spins at its corners, 
the non-shared sites of the two connected octahedra have two possibilities: 
the same or different colors in one direction. 
The \textbf{\textit{reo}} net is apt to accommodate the fluctuation 
by this freedom,\cite{reo_HeisenbergAF}
similar to the kagom\'{e} lattice\cite{syozi, 3afPotts_kagome} 
containing triangles and hexagons (in dimension two).

Lattices containing triangles have garnered significant attention
because of their ``frustrated'' nature.\cite{frustration0,frustration2,spinliquid,reo_HeisenbergAF}
In this context, it is important to remember that the term frustration 
implicitly yet essentially assumes the Ising mechanism,
which relies on two states for each spin.
Tripartite lattices are free from frustration under the assumption of 
the three states discussed in this study, 
as long as the interaction works only between the nearest neighbors.
On the other hand, at least four states are necessary to avoid frustration
in lattices containing tetrahedra, such as a face-centered cubic (FCC) lattice.
Additionally, it is important to understand the relationship 
between frustration and the degeneracy of the ground states.
While frustration generally leads to notable degeneracy,
the reverse is not universally true.\cite{exMS_2d,exMS_4d,exMS}
For instance,
any models assuming three states for each spin on bipartite lattices
exhibit notable degeneracy that results in a macroscopic entropy.\cite{MF_Potts,ono}
 
The Potts model\cite{potts,3Potts} is a generalization of 
the Ising model\cite{ising1,ising2} with the two states (up and down) for a spin
concerning the number of available states. 
The three-state Potts model can be expressed as
\begin{equation}
\frac{E}{N} = -J \sum_{\langle m,n \rangle} \bm{s}_m\cdot \bm{s}_n
\label{model}
\end{equation}
using three orthogonal spins $(1,0,0)$, $(0,1,0)$, or $(0,0,1)$ to denote three states. 
The summation in Eq.\ \ref{model} runs on interacting spin pairs 
$\langle m,n \rangle$.
The sign of $J$ determines whether the model is ferroic ($J > 0$) or antiferroic ($J < 0$). 
In the latter case, the minimum energy is 0. 
Taking into account the presence of three equivalent sublattices, 
we can perform a mean-field (Bragg-Williams) calculation\cite{MF_Potts} for the antiferroic case, 
as shown in the Appendix. 
This calculation is safe for highly symmetric cases, 
while it may be unsuitable for the present \textbf{\textit{reo}} net.
 
This paper is organized as follows. 
Section II describes the simulations for the antiferroic three-state Potts model 
on the \textbf{\textit{reo}} net. 
Section III gives proof of the plausible symmetry-breaking 
upon the self-assembly of isotropic junctions. 
The results of the simulations and discussion about them will be in Section IV. 
Section V summarizes and concludes the paper. 
Appendix describes the mean-field calculation of the model on the tripartite lattice.

\section{Simulation and Quantities of Interest}

We adopt the presentation of the antiferroic three-state Potts model of Eq.\ \ref{model}
on vertices of  the \textbf{\textit{reo}} net.
The interaction works only for connected spin pairs.

Monte Carlo (MC) simulations were performed using the Metropolis algorithm.\cite{mc} 
A trial orientation of a spin was randomly and uniformly generated
by using the Mersenne twister algorithm.\cite{MT}
At each temperature,
we performed $2\times 10^5$ steps per spin.
After the equilibration steps, 
the last quarter was used to calculate physical quantities.
The examined system sizes were $3L^3$ with $L=10$, $20$, $30$, and $50$.
Since the results were consistent,
we describe the results of $L=50$ in this paper.
For the phase transitions, thermal hysteresis was significant,
accompanying the scarce effect of fluctuations.
The normal and reverse ``quenching'' experiments were performed for $L=50$.

To see the order brought by a phase transition,
we need appropriate physical quantities,
which we can regard as kinds of the so-called order parameter(s).
In this paper, the average over a whole or part of the ensemble is denoted by 
$\langle \cdot \rangle$.

Since the ordering in the present model is mostly complete in an octahedron 
contained in a unit cell (including its boundaries),
we can use the population of the octahedron with minimum (null) energy
to express the local comfort\cite{exMS_2d} of an octahedron, equivalently a unit cell, specified by $hkl$;
\begin{equation}
c_{hkl}=
\begin{cases}
1 & \text{for $\epsilon_{hkl}=0$,}\\
0 & \text{for $\epsilon_{hkl}>0$,}
\end{cases}
\end{equation}
with 
\begin{equation}
\epsilon_{hkl}=-J \sum_{\langle m,n\rangle'} \bm{s}_m\cdot \bm{s}_n
\end{equation}
where the summation is inside the octahedron in the cell 
specified by $hkl$.
The average over the system, $c=\langle c_{hkl}\rangle$,
reflects the population of the stable, i.e., null energy octahedrons.
If $c$ is sufficiently high, the system can be regarded 
as an ensemble of the sixfold branched junctions.
Note that $c$ reflects the order but does not serve as the literal order parameter
that is finite in an ordered state and vanishes in the disordered phase.
Since this quantity requires instantaneous spin configurations,
it is estimated from 100 snapshots equally sampled in the equilibrated periods 
($5\cdot10^4$ MC steps) to reduce the computational demands.

\begin{table*}
\caption{\label{table}States observed in the simulation results of
the antiferroic three-state Potts model on the \textit{\textbf{reo}} net.}
\begin{ruledtabular}
\begin{tabular}{clll}
state (phase) & property\footnote{Assuming $\sigma_3\ge\sigma_2\ge\sigma_1$.} & 
additional properties\footnote{See Fig.\ \ref{map} 
for the definition of $a$ and $l_\mathrm{ave}$.} & 
notes\footnote{The mean-field (MF) calculation is given in Appendix.}\\
\hline
disordered & $\sigma_1=\sigma_2=\sigma_3=0\hspace{3mm}$ & $\xi=0$, $a=0$ & 
subject to the MF treatment\\
isotropic & $\sigma_1=\sigma_2=\sigma_3>0$ & $\xi=0$,  $a/l_\mathrm{ave}=\sqrt[4]{3}/2$& 
metastable, subject to the MF treatment\\ 
tetragonal & $\sigma_3>\sigma_1=\sigma_2>0$ & $\xi>0$, $a=0$ & metastable\\
 & $\langle\bm{s}_1\rangle=\langle\bm{s}_2\rangle$ &  &\\
orthorhombic\hspace{3mm} & $\sigma_3>\sigma_1=\sigma_2>0$& $\xi>0$, $a>0$ &
``typical'' ordered state\\
& $\langle\bm{s}_1\rangle\ne\langle\bm{s}_2\rangle$&  &\\
``glassy'' & $\sigma_3>\sigma_2>\sigma_1>0$ & $\xi>0$, $a>0$ &
``exceptional'' ordered state(s)\\
 \end{tabular}
 \end{ruledtabular}
 \end{table*}

Other quantities we use are based on the average spin vector on each sublattice $i$ ($i = 1,2,3$),
\begin{equation}
\langle \bm{s}_i\rangle =(\langle x_i\rangle,\langle y_i\rangle,\langle z_i\rangle).
\end{equation}
Hereafter, throughout this paper, the subscripted indices always refer to the sublattice
unless otherwise stated explicitly.
The three components are the probabilities of three states.
In the description of the simulation results,
we reorder the sublattices and states so that
$\langle z_3\rangle$ is the maximum among all components of $\langle \bm{s}_i\rangle$.
We discriminate against neither the other components nor sublattices
because they are physically equivalent in typical cases, discussed later.
Obviously, the following holds:
\begin{equation}
\langle x_i\rangle+\langle y_i\rangle+\langle z_i\rangle=1.
\end{equation}
The case with $\langle \alpha_i\rangle=1/3$ for all $\alpha$ ($=x, y, z$) on all sublattices
is the disordered state of this model.
We can imagine the state with $\langle x_1\rangle=\langle y_2\rangle=\langle z_3\rangle=1$
as that of the highest order.
However, this state is plausibly unreachable from the disordered state
because 
\begin{eqnarray}
\label{min_e_relation}
\langle z_3\rangle&=&\langle x_1\rangle+\langle x_2\rangle\nonumber\\
&=&\langle y_1\rangle+\langle y_2\rangle\\
&=&1\nonumber
\end{eqnarray}
covers
many states with the lowest, i.e., null energy,
as discussed in the next section.

We can sense the order of each sublattice by 
\begin{eqnarray}
\sigma_i &=&\sqrt{P_2(|\langle\bm{s}_i \rangle|)}\nonumber\\
&=& \sqrt{\frac{3\langle \bm{s}_i\rangle\cdot\langle \bm{s}_i\rangle -1}{2}},
\end{eqnarray}
where $P_2(\cdot)$ is the second-order Legendre polynomial.
Although its square can be used,
this makes it easier to detail in the vicinity of $1/4$. 
Obviously, $\sigma_i =0$ for the randomly occupied state,
whereas it yields $\sigma_i=1$ 
when a single state perfectly occupies the sublattice.
On the other hand, for states with the minimum energy of zero,
those of different sublattices from the unique one are
\begin{equation}
\frac{1}{2}\le\sigma_{j\ne 3}\le 1.
\end{equation}
The state with $\sigma_{j\ne 3}=1$ retains the equivalence of three sublattices,
while that with $\sigma_{j\ne 3}=1/2$ results in the tetragonal symmetry 
with the equivalence of two sublattices.
Three sublattices, hence three axes, are not equivalent in between.
In reality,
a partial order with $\sum \sigma_i\ne 0$ and $\prod \sigma_i=0$ did not happen
in the results of the present MC simulations.
That is, the former always accompanies finite $\sigma_i$ for all $i$.
Therefore, explicit descriptions about $\sigma_i$ are omitted throughout this paper.

Since the sites of each sublattice sit on different faces of a unit cell cube,
the uniaxial and isotropic self-assemblies can be judged simply by
whether the orders of the sublattices are physically equivalent or not.
To sense the non-equivalence,
we use the difference between the arithmetic and geometric means
of the norms of the average spins
$|\langle \bm{s}_i\rangle |$ of the sublattices:
\begin{equation}
\xi = \frac{1}{3}\sum_i |\langle \bm{s}_i\rangle |
-\left[\prod_i |\langle \bm{s}_i\rangle |\right]^{1/3},
\end{equation}
which is positively finite
unless they are equivalent (with vanishing $\xi$).

Table \ref{table} summarizes the properties of the five states (phases) 
encountered in the present MC simulations, 
using the quantities introduced in this section
under the assumption that $\langle z_3\rangle$ is the maximum component of
$\langle\bm{s}_i\rangle$.
Note that the $\sigma_i$s are insufficient to distinguish between 
the tetragonal and orthorhombic phases.
For this purpose, we will introduce another quantity later.

\section{Ground State and Preference into Uniaxial Order}

\begin{figure}
\begin{center}
\includegraphics[width=5.5cm]{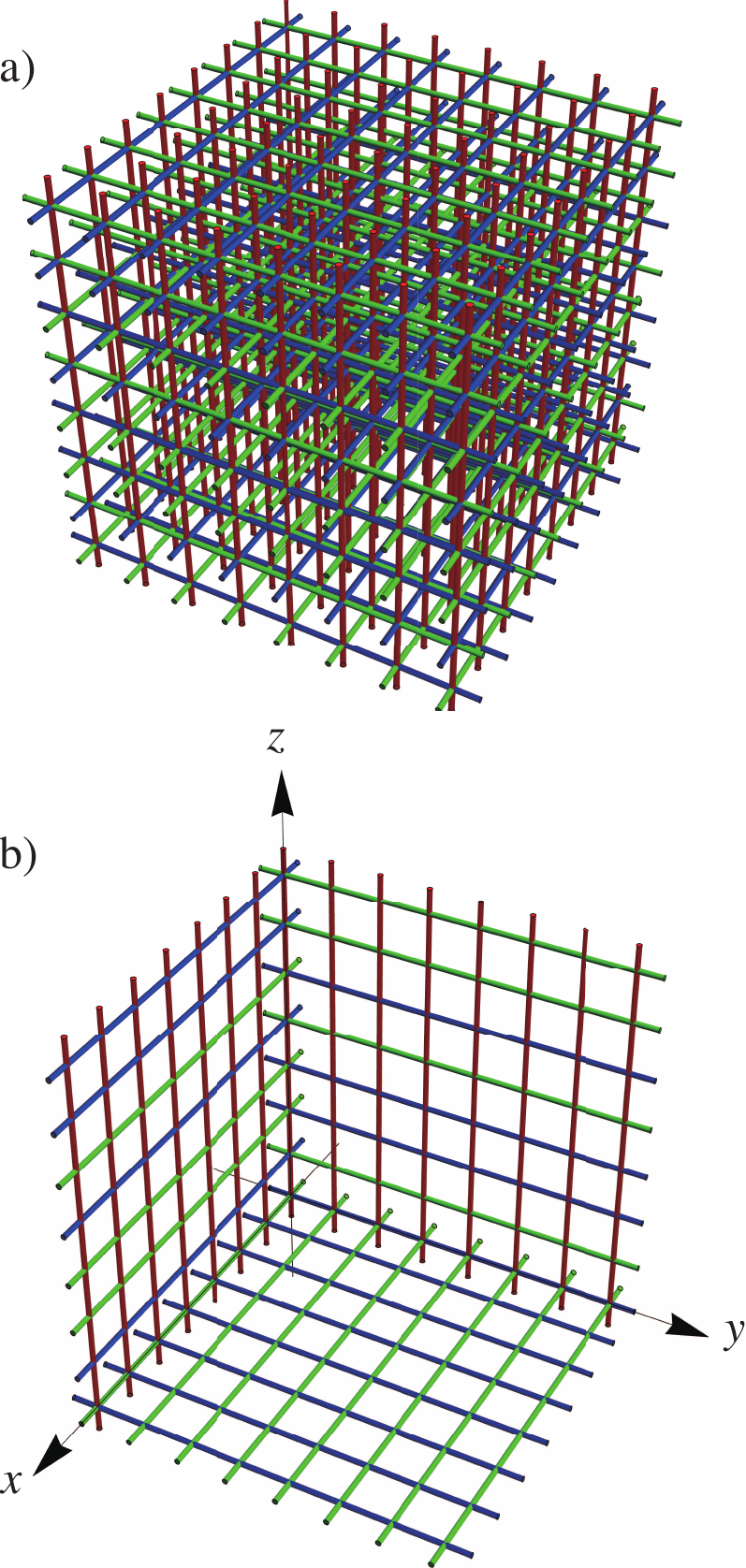}
\end{center}
\caption{a) An example of fully self-assembled states of 
$8^3$ octahedral junctions
($\langle z_3\rangle=1$, $\langle x_1\rangle=\langle y_1\rangle=
\langle x_2\rangle=\langle y_2\rangle=1/2$, and null for others),
and b) its part on the $xy$- ($z=0$), $yz$- ($x=0$), and $zx$- ($y=0$) planes.
}
\label{junglegym}
\end{figure}

We examine the variety of fully self-assembled states 
corresponding to the minimum energy of our spin model. 
Suppose that the segments of the junctions are along one of three orthogonal axes 
($x$-, $y$-, and $z$-axes). 
In fully connected states, 
the segments form a straight string of a single color 
passing through the centers of the connected junctions,
as exemplified in Fig.\ \ref{junglegym}a. 
Thus, any of the assembled states consists of flat nets (such as one on the $xy$-plane), 
which are parallel to one another. 
Each net contains strings of at least two colors 
because crossing strings (at a junction) form a right angle. 
Therefore, the fact that any single string crosses all non-parallel strings within the net 
allows only two cases: 
All strings in each of the two groups of parallel strings share the same color, 
or those in one group are mixed. 

To facilitate discussion, we place the assembly in Fig.\ \ref{junglegym}a, 
aligning it with three planes that each contain two axes, as shown in Fig.\ \ref{junglegym}b,
where only three nets on the planes are depicted.
We observe a net with two colors (G and B) on the $xy$-plane.
In this net, all strings parallel to the $x$-axis are G, while those parallel to the $y$-axis are B.
Consequently, all strings parallel to the $z$-axis must be R.
Then, the parallel nets to the basal plane consist solely of strings of G and B. 
Clearly, stacking identical layers is possible, leading to
an ideal self-assembly with the same symmetry as a single junction. 
However, there is another possibility, 
as exemplified by the second layer from the bottom.
Stacking non-rotated layers (identical to the basal plane) and 
rotated layers (as exemplified by the second layer)  
in any sequence completes the self-assembly. 
In such arrangements of any stacking orders, 
the self-assembled order is uniaxial, featuring a unique $z$-axis of the perfect order. 
The possible variety of stacking, except for the multiplicity of $6$ by color permutations, 
is enumerated as $2^L$, 
where $L$ is the number of layers.
Remarkably, 
a single layer, such as one on the $yz$- or $zx$-plane in Fig.\ \ref{junglegym}b,
is sufficient to specify the entire assembly depicted in Fig.\ \ref{junglegym}a.

In the first paragraph of this section, 
we have pointed out two possibilities regarding the number of colors in a plane net.
The case of the two colors was discussed in the previous paragraph.
Concerning the case of three colors,
we observe such layers in Fig.\ \ref{junglegym}b.
As mentioned above,
one of these layers is enough to specify the entire assembly
because all layers parallel to it must be identical.
Therefore, the counting of self-assembled states is already complete.
In this case, the enumeration is given by the serial order of B and G strings on an R string
parallel to the $z$-axis.

The number of ways to self-assemble attains a maximum
when two non-unique mixed-colored axes are equivalent 
($\langle\bm{s}_1\rangle=\langle\bm{s}_2\rangle=(1/2,1/2,0)$),
yielding the tetragonal symmetry after averaging out microscopic details.
Therefore, at this stage,
we anticipate it to be the most probable state of complete self-assembly.
Remarkably, the degeneracy at the tetragonal order does not lead to a macroscopic residual entropy because of
\begin{equation}
\lim_{L\rightarrow \infty}\frac{k_\mathrm{B} \ln 2^L}{L^3} = 0
\end{equation}
in the thermodynamic limit. 
However, the statistical weight of the intuitively straightforward self-assembly 
(with isotropic symmetry) is merely $1$ compared to $2^L$.

\section{Simulation Results and Discussion}

\subsection{Self-assembly of junctions}
In the states with the minimum energy of our antiferroic three-state Potts model, 
each diagonal spin pair on an octahedron shares the same state, 
with a different kind for each. 
Since the \textbf{\textit{reo}} lattice consists of octahedrons sharing each vertex 
with the adjacent ones, 
all spins on each array (string) along each axis share the same state. 
Each octahedron can be regarded as a junction of three strings carrying different spin states.

\begin{figure}
\begin{center}
\includegraphics[width=8cm]{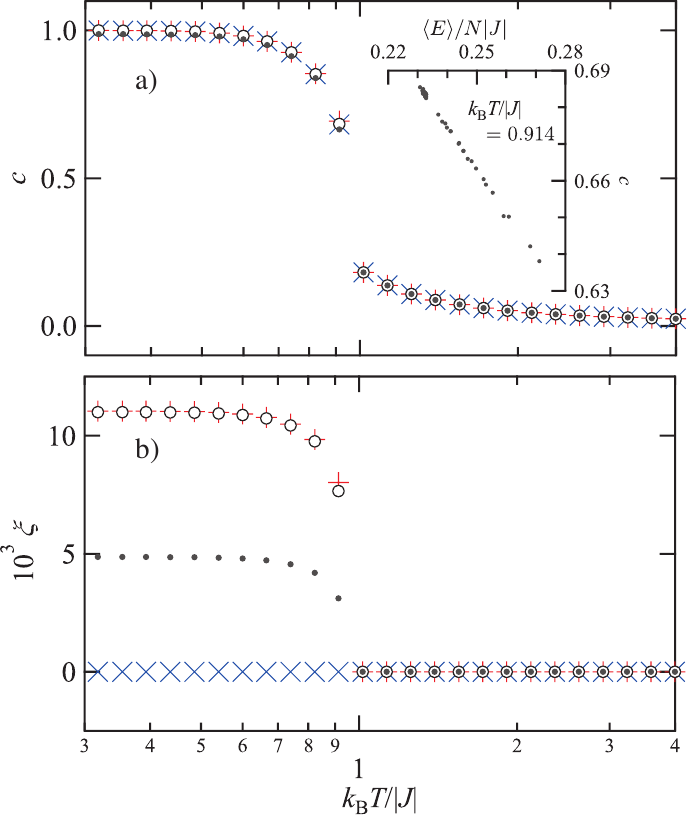}
\end{center}
\caption{a) Population ($c$) of the locally stable octahedrons ($c_{hkl}=1$) and
b) equivalence of the three sublattices ($\xi$) 
of the typical (open circle), exceptional (black dot), and two limiting (red plus and blue cross) cases
obtained for the $L=50$ system
as a function of temperature.
As the temperature scale is logarithmic,
the numerals left of $1$ are smaller than $1$, e.g., $9$ meaning $0.9$.
The typical and exceptional cases (open circle) are for the cooling from the disordered phase. 
Two limiting cases are for the heating from artificially prepared ordered states with null energy
(red plus, tetragonally ordered; blue cross, isotropically ordered).
The inset of a) is a correlation between $c$ and averaged energy at $k_\mathrm{B}T=0.914|J|$.}
\label{p_xi}
\end{figure}

We first ensure that the present MC simulations of the spin model fit the study
based on $c$, the population of stable (null-energy) octahedrons.
Figure \ref{p_xi}a shows the temperature dependence of $c$
in the entire temperature range.
A distinct and abrupt increase exists between $k_\mathrm{B}T \approx 0.9|J|$ and $|J|$.
This increase indicates that the system undergoes a phase transition.
Indeed, all $\sigma_i$ get finite below the temperature of this increase.
Since the magnitudes below and above the transition temperature lie
above and below the percolation threshold of the site process
on the simple cubic lattice \cite{SCpercolation} ($\approx 0.31160766(15)$),
the transition upon cooling can be regarded
as the occurrence of self-assembly as intended.
For the low-temperature phase,
we can consider it as an ensemble of interconnected sixfold-branched junctions
containing some defects.
Since the fluctuation, or the trend to order, in the disordered phase is insignificant
as seen in the heat capacity (see Fig.\ \ref{hysteresis}),
the phase transition to the low-temperature phase acts
as a switch of the self-assembly.
It is noted that 
the system energy $\langle E\rangle/N$ and $c$ exhibit a fairly good correspondence, 
as exemplified by the inset of Fig.\ \ref{p_xi}a for the data at $k_\mathrm{B}T = 0.914|J|$
starting from different initial conditions.
 
\begin{figure}
\begin{center}
\includegraphics[width=8cm]{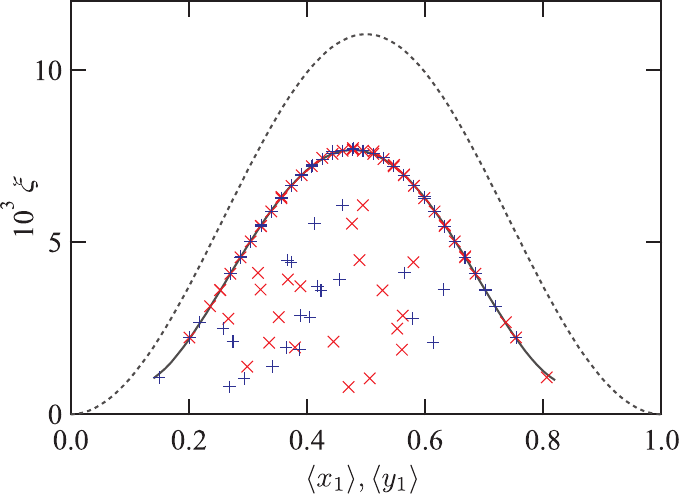}
\end{center}
\caption{Correlation between the violation of the equivalence of three sublattices ($\xi$)
and
$\langle x_1\rangle$ (crosses) or $\langle y_1\rangle$ (plus-signs) 
of the average spins ($|\langle\bm{s}_i|$) of 100 states
at $k_\mathrm{B}T = 0.914|J|$ quenched from the disordered phase. 
All points on the black solid curve correspond to 80 typical states with $\langle z_3\rangle >0.9$.
The curve remains the same when plotted against $\langle x_2\rangle$ or $\langle y_2\rangle$.
The dotted curve is $\xi$ for the null-energy spin configurations with $\langle x_1\rangle 
+ \langle y_1\rangle= \langle z_3\rangle= 1$ at $T=0$.
}
\label{xi_x}
\end{figure}

Examples of the violation of the equivalence of three sublattices are shown in Fig.\ \ref{p_xi}b.
The index $\xi$ vanishes in the high-temperature disordered phase.
In the displayed typical case,
the equivalence is broken in the low-temperature phase.
Indeed, the maximum components of average spins ($\langle z_3\rangle$)
approach unity upon cooling.
The system's comfort $c$ also approaches unity, 
implying that the system mostly reached one of the most stable spin configurations.
It was confirmed by the coincidence of the average energy and
the expected energy for non-interacting point defects, $8N|J| \exp(4J/k_\mathrm{B}T)$.
In these states of minimum energy,
we can assume Eq.\ \ref{min_e_relation},
which means $\sigma_1=\sigma_2$.
Such behaviors were confirmed in plural runs in the cooling direction.
Note that the violation $\xi$ is different from one another in typical cases.
Nearly superposed locations for a typical and the two limiting cases in Fig.\ \ref{p_xi}a are
somewhat accidental, though reflecting its distribution (see Fig.\ \ref{xi_x}).
On the other hand,  
the behavior is different in exceptional cases, as exemplified by dots in Fig.\ \ref{p_xi}.
Their energy is higher, and $c$ is smaller than the typical cases.
Indeed, they exhibit $\sigma_3>\sigma_2>\sigma_1$ 
and fail to approach zero in low temperatures.
Such failed states can be regarded as ``glassy.''\cite{glass}

\begin{figure}
\begin{center}
\includegraphics[width=8cm]{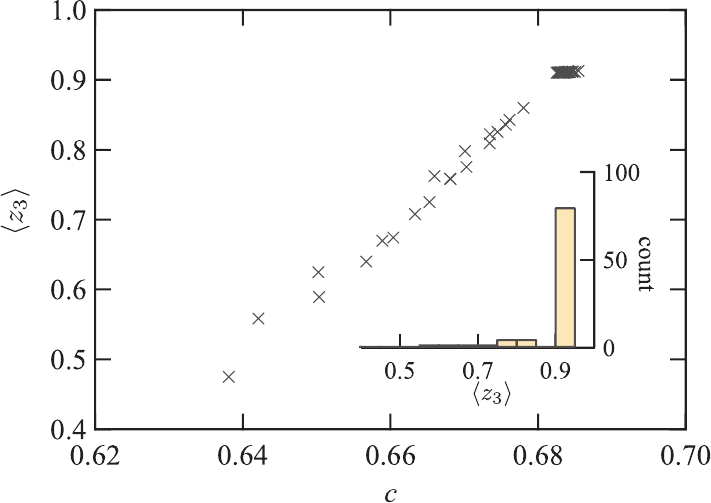}
\end{center}
\caption{Correlation between the population of null-energy octahedrons ($c$) and
the maximum component ($\langle z_3\rangle$) of the average spins ($\langle\bm{s}_i\rangle$)
of 100 states
at $k_\mathrm{B}T = 0.914|J|$ quenched from the disordered phase.
The inset shows the histogram against $\langle z_3\rangle$.
80 states have $\langle z_3\rangle>0.9$.}
\label{z_population}
\end{figure}

The ratios of $\langle x_1\rangle$ and $\langle x_2\rangle$ 
obtained in plural runs were not $1:1$
and differed significantly from one another,
in contrast to a naive expectation based on the maximization of the mixing entropy.
To get an insight into the order brought by random assembly,
we attempted 100 ``quenching'' experiments.
Simulations at $k_\mathrm{B}T = 0.914|J|$ were performed
starting from 100 snapshots during the equilibrated period at $k_\mathrm{B}T = 1.016|J|$,
where the system is in the disordered phase. 
Figure \ref{z_population} shows the correlation between $c$ and
$\langle z_3\rangle$ of 100 states obtained by quenching experiments.
An apparent positive slope indicates that smaller $\langle z_3\rangle$ implies
the failure to attain smaller energy.
There are many points with $\langle z_3\rangle>0.9$ and $c>0.68$.
Indeed, the cluster consists of 80 results from all 100 quenching experiments,
as indicated by the inset histogram. 
Therefore, the cluster represents typical cases.
Because of $\langle z_3\rangle>0.9$,
ordering along the $z$-axis is almost perfect in typical cases.
Since they all have finite $\xi$ (Fig.\ \ref{xi_x}), 
they represent the expected characteristic.

\begin{figure}
\begin{center}
\includegraphics[width=8cm]{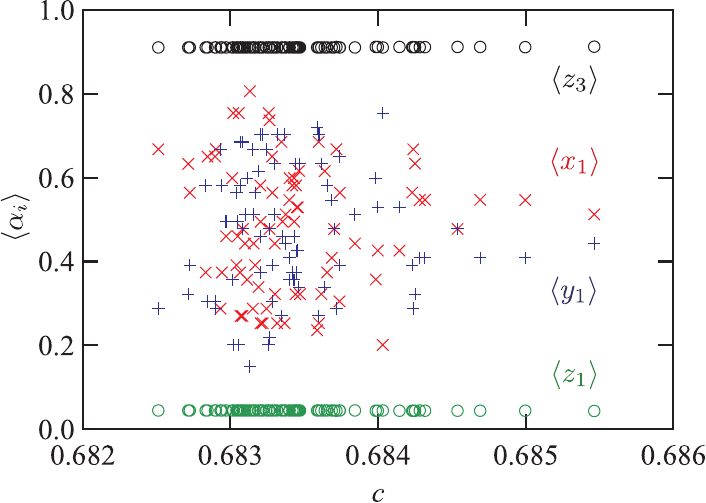}
\end{center}
\caption{Components of the average spins ($\langle\bm{s}_i\rangle$)
of 80 typical states ($\langle z_3\rangle>0.9$)
at $k_\mathrm{B}T = 0.914|J|$ quenched from the disordered phase.
The standard deviations (in the estimation from 100 snapshots) of $c$ 
is ca.\ $3\cdot10^{-3}$ for them.}
\label{quench_subL}
\end{figure}

Figure \ref{quench_subL} clarifies the details of the order in 80 typical cases.
A similar plot about the sublattice 2 is practically the same as Fig.\ \ref{quench_subL}
due to their physical equivalence 
($\langle x_1\rangle \approx \langle y_2\rangle$ and 
$\langle y_1\rangle \approx \langle x_2\rangle$).
Note that the standard deviation of $c$ in computation from 100 snapshots is ca.\ $3\cdot 10^{-3}$,
comparable to the full width of the drawn horizontal axis.
However, the magnitude of each component does not fluctuate like the plot 
from one snapshot to another; it stays mostly steady
because temperature-dependent simulation runs exhibit only moderate variation.
$\langle z_1\rangle$ is almost constant with $\langle z_1\rangle\approx (1-\langle z_3\rangle)/2$
because $\langle z_1\rangle$ and $\langle z_2\rangle$ are equally kicked out by $\langle z_3\rangle$.
Other components are distributed while keeping 
$\langle x_{i\ne 3}\rangle+\langle y_{i\ne 3}\rangle$ at an almost constant magnitude 
determined by $\langle z_3\rangle$.

The equivalence of the three sublattices was broken in
the ordered states brought by quenching, as shown in Fig.\ \ref{xi_x}.
Further, except for only a single result,
typical cases also violated the equivalence between the two sublattices ($i=1,2$) distinguished from
the maximum component $\langle z_3\rangle$.
However, all points belonging to typical cases lie on a single curve.
This curve resembles the one expected for the ensemble of null energy states:
\begin{eqnarray}
\xi &=& \frac{1}{3}\left[1+2\sqrt{\langle x_1\rangle^2+\left(1-\langle x_1\rangle\right)^2}\right]\nonumber\\
&&-\left[\langle x_1\rangle^2+\left(1-\langle x_1\rangle\right)^2\right]^{1/3}.
\end{eqnarray}
Although we do not have a theoretical curve at finite temperatures,
the resemblance suggests a similar situation.

\begin{figure}
\begin{center}
\includegraphics[width=8cm]{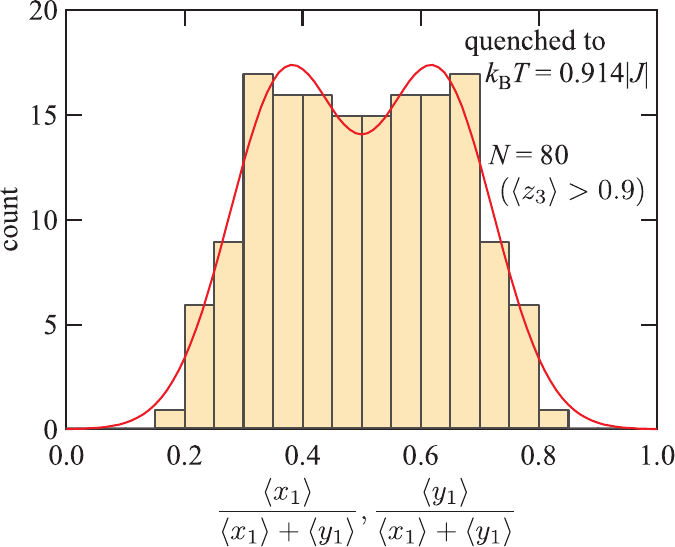}
\end{center}
\caption{Histogram of the relative asymmetry of $\langle x_1\rangle$ and $\langle y_1\rangle$
of 80 typical states ($\langle x_1\rangle+\langle y_1\rangle\approx \langle z_3\rangle>0.9$)
at $k_\mathrm{B}T = 0.914|J|$ quenched from the disordered phase.
The normalization by $\langle x_1\rangle+\langle y_1\rangle$ corrects
the effect of $\langle z_1\rangle\ne 0$ at finite temperatures.
The histogram concerning the sublattice 2 is practically the same.
The total counts of the histogram are $2\times 80$ due to doubled counts against
$\langle x_1\rangle$ and $\langle y_1\rangle$.
The solid continuous curve is a fit result by the sum of two normal distribution functions equally 
displaced from $1/2$.}
\label{double_peak}
\end{figure}

\begin{figure}
\begin{center}
\includegraphics[width=8cm]{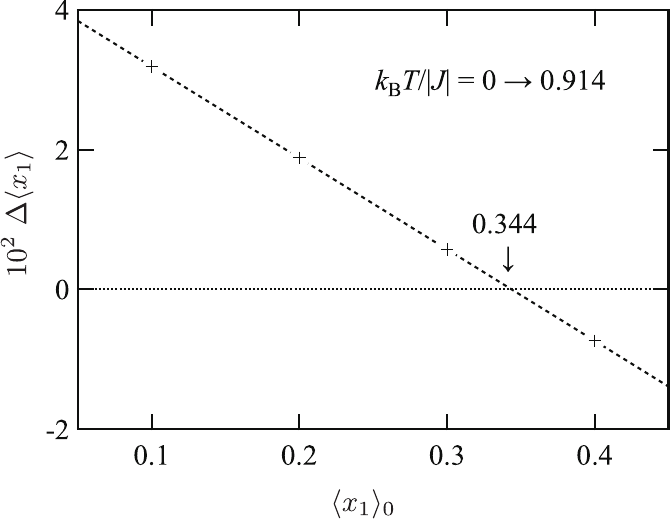}
\end{center}
\caption{Shifts $\Delta\langle x_1\rangle=\langle x_1\rangle-\langle x_1\rangle_0$ 
in the reverse quenching to $k_\mathrm{B}T/|J| = 0.914$ from
the respective initial conditions corresponding to $T=0$
($\langle x_1\rangle_0=0.1, 0.2, 0.3$, and $0.4$).
}
\label{deltaX0}
\end{figure}

The fluctuation in $c$ up to ca.\ $3\cdot 10^{-3}$ implies that
the histogram of relative ratios between $\langle x_1\rangle$ and $\langle y_1\rangle$ 
should be meaningful.
Figure \ref{double_peak} shows a symmetrized histogram against the relative magnitudes
for 80 typical states at $k_\mathrm{B}T = 0.914|J|$ quenched from the disordered phase.
A remarkable feature is its shape.
It is neither singly peaked at $1/2$ as expected from the simple mixing rule,
nor flat, but doubly peaked.
The plausibility of this shape is supported by the results of the reverse ``quenching,'' 
i.e., temperature jump experiments:
MC simulation runs set at $k_\mathrm{B}T = 0.914|J|$ starting from the initial spin configurations 
of the randomly prepared null energy states (corresponding to $T=0$) with
$\langle x_1\rangle_0=0.1, 0.2, 0.3$,  and $0.4$ yield $\langle x_1\rangle$ different magnitudes
from the initial, as shown in Fig. \ref{deltaX0}.
Remarkably, the shift $\Delta\langle x_1\rangle=\langle x_1\rangle-\langle x_1\rangle_0$ 
varies linearly and 
changes its sign between $\langle x_1\rangle_0=0.3$ and $0.4$,
yielding the fixed point of  $\langle x_1\rangle_\mathrm{fix}=0.344$.
This fixed point corresponds to $0.36$ in the reduced scale in Fig.\ \ref{double_peak}  and
is reasonably close to the peak of the histogram.
The linear dependence of $\Delta\langle x_1\rangle$ can be interpreted 
as a result of the initial ``impact'' proportional to the deviation from the fixed point
under an overdamped situation.

Although we imagine the tetragonal (uniaxial) order 
with $\langle\bm{s}_1\rangle=\langle\bm{s}_2\rangle$ for the ordered phase 
because of the maximizing the mixing entropy of the two orientations of layers, 
the two axes within a layer seem to be inequivalent in the simulation results, 
implying the unequal number of the two orientations. 
If we consider the stack of the perfectly ordered layers, 
they are classified into two orientations. 
Maximizing the entropy yields an equal number of two types of layers, as discussed before. 
The attainable entropy is $Lk_\mathrm{B} \ln 2$. 
If the dependence of the energy on $\xi$ is not convex upward,
the number of the two orientations should be the same.
The dependence is flat at absolute zero.
If the dependence is convex upward,
the possible appearance of the two maxima in probability may happen
depending on the degree of the convexity.
In this case, the effectiveness of this mechanism should rely on the system size,
and it would disappear in the thermodynamic limit, leading to the first expectation in the limit.
However, the necessary dependence was not observed 
in the energy of the quenched states (not shown).
Therefore, this mechanism seems to be inapplicable in the present case,
assuming successful sampling of states in only 100 quenching trials
of the ensemble of a third of a million.

On the other hand, 
layers can be distinguishable from each other at finite temperatures 
by imperfections of the order within each layer. 
A naive combinatorial estimate of the entropy yields 
$k_\mathrm{B} \ln L! \approx Lk_\mathrm{B}(\ln L -1)$, 
which is significantly larger than, though still non-macroscopically, 
the case of two orientations of perfect layers. 
Thus, the doubled peak may have a physical origin. 
In this case, we cannot assume a smooth convergence with decreasing temperature 
to the most stable state with  $\langle x_1\rangle=1/2$ at absolute zero.

The choice of how to stack stable layers resembles the case of rigid spheres.\cite{HS1,HS2,HS3,HS4,HS5,HS6,HS7}
However, there is a difference from the present case.
While the ensemble of rigid spheres is solely governed by entropy,
the present case suffers from both entropy and energy.
Furthermore,
it is hard to imagine any other protocol than the gradual compression (or inflation of the sphere radii)
to see the process of the stacking order selection in the rigid sphere case.
This protocol seems to be qualitatively different from the present case.
Still, the similarity seems to deserve further consideration in the future.

\subsection{Three Ordered Phases and Relative Phase Stability}

\begin{figure}
\begin{center}
\includegraphics[width=8.2cm]{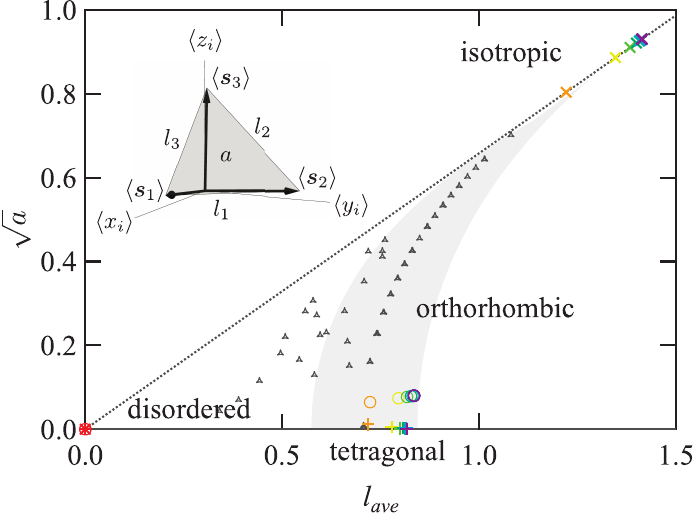}
\end{center}
\caption{Phase map of the antiferroic three-state Potts model on the \textbf{\textit{reo}} net,
a balanced tripartite lattice.
The coordinates are the average edge length ($l_\mathrm{ave}\ =(l_1+l_2+l_3)/3$) 
and the square root of the area ($\sqrt{a}$)
of the triangle formed by three points corresponding to the populations of the respective spin states
on three sublattices ($\langle\bm{s}_i\rangle$, $i=1,2,3$),
as shown in the inset.
In general, a point further from the origin in each phase corresponds to a lower temperature.
A probable domain of the orthorhombic phase is shaded.
The isotropic and tetragonal phases are the two parts of the periphery of the orthorhombic region.
The isotropic phase lies on the dotted line (the relation for equilateral triangles) except for the origin,
which corresponds to the disordered phase.
The tetragonal phase is on the horizontal null axis.
The data obtained in the MC simulations covering a series of temperatures
are plotted in rainbow colors by different symbols as examples:
cross, isotropic phase $\rightarrow$ disordered phase (heating);
plus, tetragonal phase $\rightarrow$ disordered phase (heating);
circle, disordered phase $\rightarrow$ orthorhombic phase (cooling).
Small triangles indicate the results of the quenching to $k_\mathrm{B}T=0.914|J|$
from the disordered phase at $k_\mathrm{B}T=1.016|J|$.
}
\label{map}
\end{figure}

The simulation results of normal and reverse quenching runs described in the previous section
imply that states with $0<\langle x_1\rangle<1/2$ are typical at finite temperatures.
However, the states with $\xi=0$ were found to be stable
except for a close vicinity of the upper bound of phase stability
once the simulation runs were started from the null energy state of
$\langle x_1\rangle=\langle y_2\rangle=\langle z_3\rangle=1$.
It is also the case for
the state of $\langle\bm{s}_1\rangle=\langle\bm{s}_2\rangle$
($\langle x_1\rangle=\langle x_2\rangle\approx \langle z_3\rangle/2$) starting from
$\langle x_1\rangle=\langle x_2\rangle=\langle z_3\rangle/2=1/2$ with random stacking of
null energy nets parallel to the $xy$-planes.
This state was also obtained once (in $100$ trials)
in the quenching experiment from the disordered phase.
They have a different symmetry in lattice geometry from the typical cases
discussed in the previous section.
The state of the equivalence of three axes ($\sigma_1=\sigma_2=\sigma_3$) 
can be characterized as isotropic,
assuming the physical equivalence of three spin states and
the cubic symmetry of the underlying \textbf{\textit{reo}} lattice.
The strict equivalence of two axes with a unique axis 
($\langle\bm{s}_1\rangle=\langle\bm{s}_2\rangle\ne\langle\bm{s}_3\rangle $) 
can be tetragonal.
Contrastingly, the typical case of $\sigma_1=\sigma_2$ with
$\langle\bm{s}_1\rangle\ne\langle\bm{s}_2\rangle$
 ($\langle x_1\rangle=\langle y_2\rangle\ne\langle z_3\rangle$) can be said to be orthorhombic.
Therefore, they should be regarded as distinct phases.
Indeed, they, including the disordered phase,
clearly belong to different domains in a space spanned by
the populations of spin states (colors).
We consider triangles with their vertices 
at
$\langle \bm{s}_1\rangle$, $\langle \bm{s}_2\rangle$, and $\langle \bm{s}_3\rangle$
belonging to an equilibrium state, as depicted in the inset of Fig.\ \ref{map}.
Plotting the square root of the area $a$
(for the homogeneous dimension of the quantities)
against the mean length of the edges $l_\mathrm{ave}$
yields the main panel of  Fig.\ \ref{map}.
This map indicates the relationship among phases.
The disordered phase shrinks to a point (origin).
In contrast, the orthorhombic phase has some area,
of which the two parts in its periphery correspond to 
the isotropic phase (on the slanted straight line expressing equilateral triangles,
$\sqrt{a}/l=\sqrt[4]{3}/2$) and
tetragonal phase (on the horizontal axis because of 
$a=0$ after $\langle\bm{s}_1\rangle=\langle\bm{s}_2\rangle$)
due to their involvement of equality among/between the characterizing quantities
(see Table \ref{table}).
Only the isotropic phase is within the scope of mean-field treatment,
as discussed in Appendix.

Since all three phases share the null energy at absolute zero,
their free energies start from the same value.
While the isotropic phase has vanishing entropy,
the other two phases have finite entropy due to the mixing entropy of
the null energy stacking.
It is maximal in the tetragonal phase.
Therefore, the free energy at very low, but finite, temperatures should fulfill the relation:
$F_\mathrm{tetra}<F_\mathrm{ortho}<F_\mathrm{iso}$.
On the other hand, the simulation results suggest that
the orthorhombic phase is the most stable around $k_\mathrm{B}T\approx 0.9|J|$,
implying the crossing of the free energy between the tetragonal and orthorhombic phases.
However, no symptom was detected in the simulation results.
This is not surprising in two respects.
The first is that the mixing entropy under this discussion is not macroscopic,
 i.e., minimal,
 implying that the driving force of the expected phase transition is very small.
 The second is, consequently, that the variation of $\langle \alpha_i\rangle$ (of ordered phases)
 is weak in a series of simulations in heating and cooling.
 This is partly due to the high $c$, i.e., well-connected octahedrons,
 resulting in a situation where local change of the spin configuration is not easy.
 
\begin{figure}
\begin{center}
\includegraphics[width=8cm]{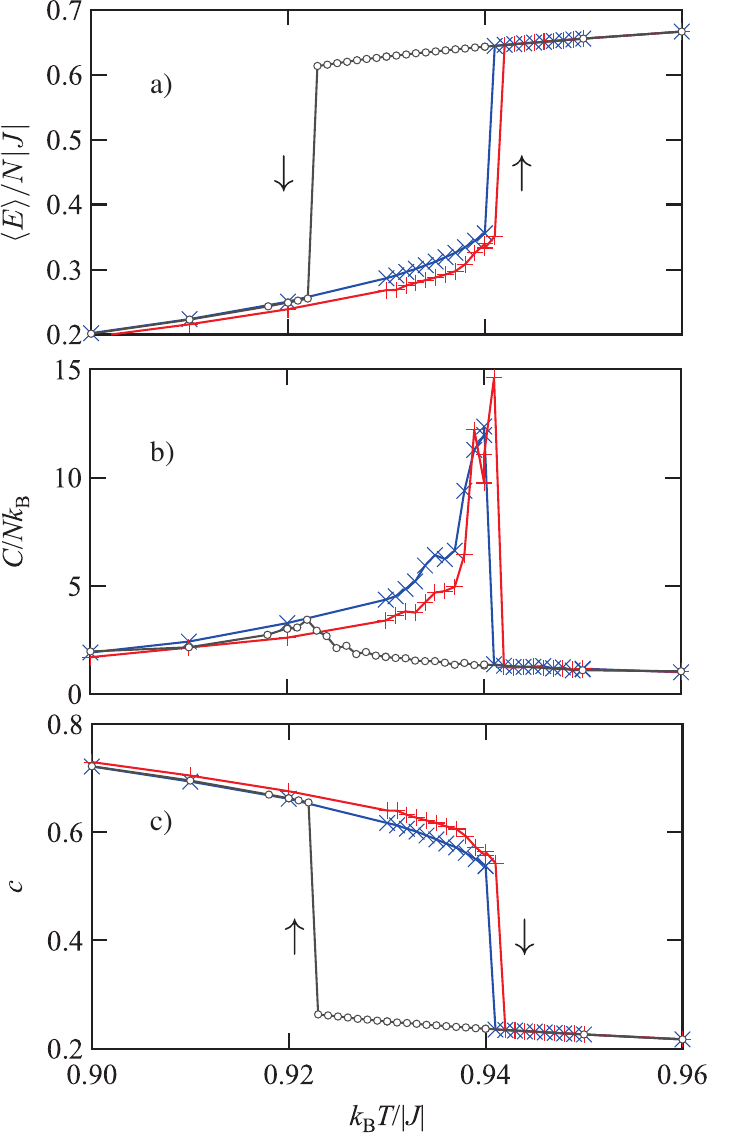}
\end{center}
\caption{Examples of MC simulation data around the phase transitions between
an ordered 
[isotropic (blue cross, on heating runs), tetragonal (red plus, on heating runs), 
or orthorhombic (open circle, on cooling)] phase and the disordered phase.
a) average energy, b) heat capacity, c) population of null-energy octahedrons.
}
\label{hysteresis}
\end{figure}

Figure\ \ref{hysteresis} shows the details around the phase transitions.
Significant hysteresis in the cooling and heating runs indicates that 
the phase transition is of the first order.
Although the isotropic and orthorhombic phases exhibit 
nearly degenerate energy and comfort $c$,
the tetragonal phase is distinct from them.
Still, their difference is not large.
We do not attempt to determine the equilibrium transition temperature precisely here.
It happens around $k_\mathrm{B}T\approx 0.93|J|$
with an entropy increment of about $\Delta_\mathrm{trs}S/N\approx \Delta E/NT
=0.35/0.93\approx\ 0.4 k_\mathrm{B}$,
about one-third of the full entropy of $k_\mathrm{B}\ln 3\approx 1.1k_\mathrm{B}$.
The location of the phase transition is lower than that of the continuous transition
reported for the simple cubic lattice
($k_\mathrm{B}T_\mathrm{c}\approx 1.23|J|$),\cite{ono}
despite the larger number of surrounding spins.
This suppression may be due to the significant fluctuation 
inherent in the structure of the \textit{\textbf{reo}} net,
although we need to consider the issue 
that the low-temperature phase of the simple cubic lattice
is a Kosterlitz-Thouless phase\cite{ono} to reach a conclusion.
The presence of a macroscopic entropy at absolute zero for the simple cubic lattice
can also be an essential difference.
Indeed, it is interesting to note that
the ratio of the transition temperatures of the present model and that of the continuous transition
on the triangular lattice ($k_\mathrm{B}T_\mathrm{c}\approx 0.63|J|$),\cite{triangularPotts}
which is also a balanced tripartite lattice,
is not far from the ratio between the numbers of interacting spins ($8/6$).

In Fig.\ \ref{hysteresis},
a kink can be recognized  at $k_\mathrm{B}T\approx 0.936|J|$ in the energy and the comfort
of the tetragonal phases.
At this temperature,
a slight deviation from the tetragonal symmetry occurs:
$\langle \bm{s}_1\rangle= \langle \bm{s}_2\rangle\rightarrow
\langle \bm{s}_1\rangle\ne \langle \bm{s}_2\rangle$.
This kink appears to accompany a peak in heat capacity.
Therefore, the kink can be a symptom of a phase transition
from the tetragonal phase to a kind of orthorhombic state (phase).

\section{Summary and Conclusion}

This paper has examined the self-assembly of the isotropic sixfold (octahedral) junctions,
paying attention to the similarity/difference in the symmetries of
the elementary entities and the self-assembled order.
The rigorous discussion shows,
counterintuitively, that the self-assembled order is not isotropic but uniaxial.
Uniaxial means that the order is complete in a single unique direction.
In contrast, the order along the two perpendicular directions can be mixed
with a non-macroscopic but large degeneracy,
leading to the expectation of the tetragonal order guided by 
the maximization of the mixing entropy.

The appearance of the uniaxial order was assessed by numerical simulations.
For this purpose, an antiferroic three-state Potts model on the \textbf{\textit{reo}} net
was adopted.
The comparison with the site percolation on the simple cubic lattice
ensured the extended self-assembly in low-temperature ordered phases.
The formation of the expected uniaxial order was established.
A possible variety of ordered phases, i.e.,
the isotropic, tetrahedral, and orthorhombic phases, and their relations
were briefly discussed, as summarized in Table \ref{table}.

The quenching experiments from the disordered phase on 100 samples 
gave a characteristic distribution of the order parameters perpendicular to
the unique axis of mostly perfect order.
The reverse quenching experiments from artificially prepared initial states of null energy
suggested that this doubly-peaked distribution was intrinsic.
Since the distribution differs from what is expected from the simple maximization 
of the mixing entropy,
some mechanism should exist.
A possible connection between this finding and the relative stability of 
a stacking variety of hard spheres
was pointed out.
We suggested the possibility of distinguishability of stacking layers at finite temperatures.

The quenching experiments also revealed
that the system is trapped in a ``glassy'' state with a non-negligible probability
at the phase transition to a low-temperature ordered phase.
The present model, an antiferroic three-state Potts model on the \textbf{\textit{reo}} net,
can be used as a model capable
of providing ordered/disordered-trapped states depending on the protocols.

A similar application of spin models seems to be applicable to various self-assemblies.
For instance, 
similarities with polynuclear complexes are evident for the current octahedral junctions. 
Therefore,  
modifying the assumed lattice geometry and/or the employed model itself 
could tune the present model in alignment 
with an issue of the formation mechanism of metal-organic frameworks 
(MOFs).\cite{mof1,mof2,mof3,mof4,mof5,mof6}
In the present application using an antiferroic three-state Potts model,
the transition from the high-temperature disordered phase
to an ordered state served as a switch of the self-assembly.
We can imagine 
a jump in temperature and/or pH or the addition of a poor solvent
as actual examples of such a switch. 
However,
the potential applicability of the method extends beyond such straightforward examples
because a lattice can be complex at will, solely reflecting the connectivity of the target
despite the intricacies of the analysis.
In this context, the antiferroic \textit{four}-state Potts model is promising\cite{cluster}
because, after the four-color theorem,\cite{4color} 
it is compatible with the vertices (spin sites) 
of any nets on the surface of a body homeomorphic to a sphere in dimension three.

Finally, a comment about our characterization
of the sixfold blanched junction as ``isotropic'' is included.
If we take the colors of the three end pairs seriously,
the symmetry of each junction is not isotropic but orthorhombic.
In this case, the self-assembly can be discussed
in the context of the so-called ``order-by-disorder''\cite{ObyD}:
Although this mechanism predicts the tetragonal order,
a minimal entropy gain allows the preservation of the orthorhombic symmetry,
i.e., non-equivalence of three axes, in many occasions of self-assembly.



%
%

%
 

\begin{acknowledgments}
The author thanks Dr.\ Y.\ Yamamura for his valuable comment
for improving this paper.

This work was supported by JSPS KAKENHI, 
Transformative Research Areas (A) ``Materials Science of Meso-Hierarchy'' No.\ 24H01694.
\end{acknowledgments}

\section*{Declarations}
The author has no conflicts to disclose.

The author played all roles concerning this study:
Conceptualization, Software, Data Curation, Formal Analysis, Funding Acquisition,
Visualization, Writing \& Editing.

\section*{Data Availability Statement}
Further data that support the findings of this study are available
 from the author upon reasonable request.

\appendix*

\section{Mean-Field Treatment of Antiferroic Three-State Potts Model 
on Balanced Tripartite Lattices}

In this Appendix, we formulate the mean-field (Bragg-Williams) analysis
of the antiferroic three-state Potts model on balanced tripartite lattices, 
following the reference,\cite{MF_Potts}
and show the inapplicability of the approximation to the uniaxial ordering discussed in the main text.
We denote the probability of the state C ($=$\ R, G or B) on the sublattice $i$ ($=1-3$)
by $p_{\mathrm{C},i}$.
Obviously, the following should hold:
\begin{equation}
1=\sum_{\mathrm C} p_{\mathrm{C},i}.
\label{sum_rule}
\end{equation}
The average energy of the model consisting of $N$ spins is
\begin{equation}
\frac{\langle E\rangle}{N}=
-\frac{zJ}{6}\sum_{\langle i,j\rangle}\sum_\mathrm{C} p_{\mathrm{C},i} p_{\mathrm{C},j},
\end{equation}
where the first summation runs over all combinations of sublattices except for  $i= j$.
The factor $6$ instead of the ordinary $2$
comes from the presence of three sublattices.
For the same reason, the entropy is given by
\begin{equation}
\frac{\langle S\rangle}{N}=- \frac{k_\mathrm{B}}{3}\sum_i\sum_\mathrm{C}
p_{\mathrm{C},i} \ln p_{\mathrm{C},i}.
\end{equation}
The minimization of the free energy, $\langle E\rangle - T\langle S\rangle$,
with respect to $p_{\mathrm{C},i}$ results in
\begin{equation}
0= -\frac{zJ}{6}\sum_{j\ne i}p_{\mathrm{C},j}
+ \frac{k_\mathrm{B}T}{3}\left(\ln p_{\mathrm{C},i}+1\right)+\lambda_i
\label{for_i}
\end{equation}
including a Lagrange multiplier $\lambda_i$ to reflect Eq.\ \ref{sum_rule}.
Summing Eq.\ \ref{for_i} over the color yields
\begin{equation}
\lambda_i=\frac{zJ}{9}-\frac{k_\mathrm{B}T}{3}\left\{1
+\ln \left(p_{\mathrm{R},i} p_{\mathrm{G},i} p_{\mathrm{B},i}\right)^{1/3}\right\}.
\end{equation}
Therefore, the equation to be solved is
\begin{equation}
\frac{zJ}{2}\left(2-3\sum_{j\ne i}p_{\mathrm{C},j}\right)
+ k_\mathrm{B}T\ln \frac{p_{\mathrm{C},i}^3}{\prod_\mathrm{C}p_{\mathrm{C},i}}=0.
\end{equation}
 
If we assume the physical equivalence of both the sublattices and the colors,
we may write as
\begin{eqnarray}
p_{\mathrm{R},1}&=& p_{\mathrm{G},2}=p_{\mathrm{B},3}\nonumber\\
p_{\mathrm{G},1}&=& p_{\mathrm{B},2}=p_{\mathrm{R},3}\\
p_{\mathrm{B},1}&=& p_{\mathrm{R},2}=p_{\mathrm{G},3}\nonumber
\end{eqnarray}
for the antiferroic case.
Then, we have
\begin{equation}
\frac{3zJ}{2}\left(p_{\mathrm{R},i}-\frac{1}{3}\right)
+k_\mathrm{B}T\ln \frac{p_{\mathrm{R},i}^2}{p_{\mathrm{G},i}p_{\mathrm{B},i}}=0
\end{equation}
for all sublattices by Eq.\ \ref{sum_rule}.
Obviously, $p_{\mathrm{C},i}=1/3$ for all sublattices and colors is the trivial solution
corresponding to the disordered phase.
For the ferroic case,
$p_{\mathrm{C},i}$ should be assumed to be common for all sublattices,
yielding the first factor of Eq.\ \ref{mf_eqn} of $-3zJ/2$ instead of $3zJ/4$.
Therefore, the ferroic case has exactly twice 
the mean-field transition temperature of the antiferroic case with the same $|J|$.

\begin{figure}[tb]
\begin{center}
\includegraphics[width=8cm]{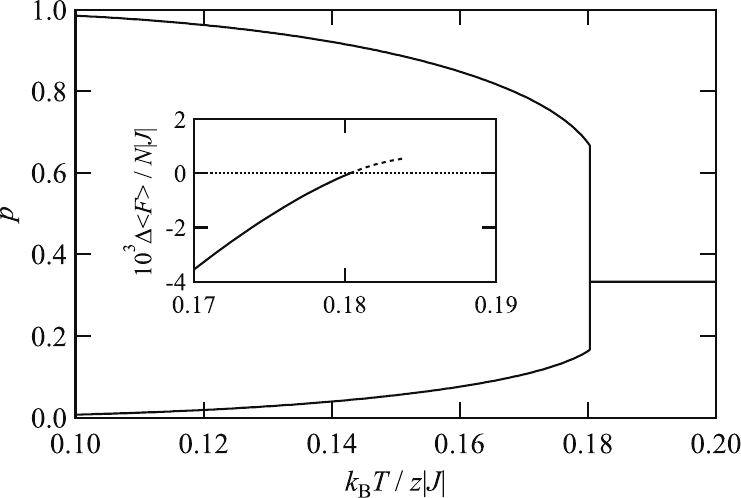}
\end{center}
\caption{Temperature dependence of 
the probabilities of the dominant ($p>1/3$) and other ($p<1/3$) states
obtained by the mean-field calculation of the antiferroic three-state Potts model
on balanced tripartite lattices.
The inset shows the sign reversal of the excess free energy beyond the disordered phase at
$k_\mathrm{B}T_\mathrm{MF}/z|J|=(8\ln 2)^{-1}\approx 0.1803$, 
at which $p_\mathrm{dominant}=2/3$.}
\label{mf_fig}
\end{figure}

If we further assume
$p_{\mathrm{R},i}\ne p_{\mathrm{G},i}=p_{\mathrm{B},i}$,
we have
\begin{equation}
p_{\mathrm{G},i}=p_{\mathrm{B},i}=\frac{1-p_{\mathrm{R},i}}{2}
\end{equation}
because of Eq.\ \ref{sum_rule}.
Then, we reach the equation of a single unknown variable,
\begin{equation}
\frac{3zJ}{4}\left(p_{\mathrm{R},i}
-\frac{1}{3}\right)+k_\mathrm{B}T\ln \frac{2p_{\mathrm{R},i}}{1-p_{\mathrm{R},i}}=0,
\label{mf_eqn}
\end{equation}
which can be solved numerically.
At low temperatures, Eq.\ \ref{mf_eqn} has three solutions,
of which the largest one is the most stable, as shown in Fig.\ \ref{mf_fig}.
This solution corresponds to the metastable isotropic case found in the simulations
described in the main text.
Because of the asymmetry for the positive and negative sides of the corresponding order parameter
$\sigma=(3p_{\mathrm{R},i}-1)/2$, the transition between this ordered state
and the disordered state ($\sigma=0$) is of first order.\cite{landau,toledano}
However, this conclusion is valid at most only a half for highly symmetric tripartite lattices
because the transition of the triangular lattice ($z=6$) is continuous
with $k_\mathrm{B}T_\mathrm{c}=(0.627163 \pm 0.000003)|J|$.\cite{triangularPotts}
A reasonable comparison may be possible 
for the icositetrachoron honeycomb in dimension four.\cite{polyhedron,exMS_4d}

A tetragonal order with the maximum entropy in self-assembly should correspond to
\begin{eqnarray}
p_{\mathrm{R},1}&\ne & p_{\mathrm{G},1}
= p_{\mathrm{B},1}=\frac{1-p_{\mathrm{R},1}}{2}\nonumber\\
p_{\mathrm{R},2}&=&p_{\mathrm{R},3}\\
p_{\mathrm{G},2}&=&p_{\mathrm{B},3}
=p_{\mathrm{B},2}=p_{\mathrm{R},3}=\frac{1-p_{\mathrm{R},2}}{2}\nonumber.
\end{eqnarray}
However, such assumptions yield exactly the same equation as Eq.\ \ref{mf_eqn},
which only gives the solution in Fig.\ \ref{mf_fig}.
This fact means that the tetragonal order is beyond the mean-field treatment
due to the essential role of local correlations.



\end{document}